\documentclass[aip,apl,a4,reprint,twoside,floatfig,superscriptaddress]{revtex4-1} 

\usepackage{graphicx}
\usepackage{amsmath}
\usepackage{amssymb}
\usepackage{amsfonts}
\usepackage{dcolumn}
\usepackage{epsfig}
\usepackage{bm}
\usepackage{calc}

\begin{document}

\title{Fabrication of Glass Micro-Cavities for Cavity QED Experiments}
\author{Arpan Roy}
\author{Murray D. Barrett}
\email{phybmd@nus.edu.sg}
\affiliation{Centre for Quantum Technologies and Department of
  Physics, National University of Singapore, 3 Science Drive 2, 117543 Singapore}

\begin{abstract}

We report a process for fabricating high quality, defect-free spherical mirror templates suitable for developing high finesse optical Fabry-Perot resonators. The process utilizes the controlled re-flow of borosilicate glass and differential pressure to produce mirrors with 0.3 nanometer surface roughness. The dimensions of the mirrors are in the $0.5-5\,\mathrm{mm}$ range making them suitable candidates for integration with on-chip neutral atom and ion experiments where enhanced interaction between atoms and photons are required. Moreover the mirror curvature, dimension and placement is readily controlled and the process can easily provide an array of such mirrors. We show that cavities constructed with these mirror templates are well suited to quantum information applications such as single photon sources and atom-photon entanglement.
\end{abstract}
\maketitle
High finesse optical cavities have been widely used to enhance the interaction between atoms and photons. Most of these experiments have been restricted to macroscopic Fabry-Perot cavities made from super polished, concave mirrors with resulting cavity finesses in the order of a million\cite{Meschede}.  The potential applications of this technology has motivated research into other fabrication methods to provide greater flexibility in experimental design.  Notable examples of this are the use of fiber based cavities \cite{Reichel}, and silicon based processing \cite{Hinds1,sandia}.  Both these approaches have been motivated by the desire to incorporate the cavity technology with atom chips \cite{Hinds2,Reichel2,Schmiedmayer1,Schmiedmayer2,Schmiedmayer3} and ion traps.  The potential of these approaches has been demonstrated \cite{Hinds2, Reichel} with each having their own unique advantages and disadvantages.  Silicon based processing relies on the sophisticated fabrication techniques that have been well developed by industry.  Cavities based on this approach can have very small dimensions which leads to a small mode volume and a very high atom-cavity coupling factor \cite{Hinds1,sandia}.  However, the small dimensions make it difficult to combine these cavities with atomic physics experiments, particularly for ion traps where anomalous heating scales strongly with the atom-to-surface distance \cite{Wineland}.  Fiber based cavities \cite{Reichel} have been successfully integrated with atom chips but integration with ion traps is still likely to be problematic.  Moreover mode coupling to the cavity is completely determined by the fiber geometry and cannot be easily controlled.  This has limited experiments to atom detection where one is not concerned with the mode in which the photon exits the cavity. Atom-photon entanglement and remote entanglement, which are of interest to quantum information processing, require the exiting photon to be in a well defined mode of the electromagnetic field.

Here we report a process for fabricating high quality spherical mirror templates using a controlled re-flow of borosilicate glass. Our process provides mirror curvatures in the $0.5-5\,\mathrm{mm}$ range. These dimensions are small enough to capitalize on the advantages of shrinking the dimensions of the cavity mirrors \cite{Isaac2,Isaac3} but large enough to avoid problems associated with surface charging and small atom-surface distances \cite{Wineland}.  In addition, our process allows us to fabricate near hemispherical mirrors suitable for enhanced light collection from a single ion\cite{Kim2,Blinov1} or neutral atom\cite{Kim1}, tight focussing of light to an atom\cite{Hinds3, Christian} and the possible suppression of spontaneous emission in the presence of a spherical reflector. \cite{Blatt}.

Fabrication methods using re-flowed borosilicate glass have been used before to create mirrors for high finesse cavities\cite{Raymer} but controlling the location and dimensions of the mirrors was problematic and mirror curvatures were typically $<200\mathrm{\mu m}$.  Instead, we utilize a modification of a method used to micro-fabricate vapour cells\cite{eklund2007glass}.  This technique used the expansion of trapped gas to blow glass into a spherical shell.  Here we use a trapped partial vacuum to deform the glass into the desired concave surface.

Blind holes approximately $2\,\mathrm{mm}$ deep are first drilled into a Macor substrate.  Macor is used as it is easily machinable and has similar mechanical and thermal expansion properties to borosilicate.  The substrate is then placed into a high temperature vacuum furnace with a $100\,\mathrm{\mu m}$ thick coverslip placed loosely over the blind holes.  The furnace is sealed and evacuated to a pressure of $300\,\mathrm{mBar}$. The temperature is then ramped at the rate of $5\,\mathrm{^\circ C/min}$ to a final temperature of $800\,\mathrm{^\circ C}$.  At this temperature the borosilicate glass softens and bonds to the Macor, securing the blind holes with a vacuum tight seal.  Air is then introduced into the furnace at a fixed pressure, typically $700\,\mathrm{mBar}$. Under these conditions, the glass behaves like a visco-elastic membrane with a circular boundary. The pressure difference and the surface tension together give rise to the spherical shape of the mirror template and the deformation process can be accurately modeled in a similar way to the treatment of micro-fabricated vapor cells. \cite{eklund2007glass}. Controlling the pressure difference and the dimensions of the cylindrical holes allows us to deterministically tune the radius of curvature of the resulting mirror substrate.

An important part of the fabrication process is returning the furnace to room temperature.  Borosilicate glass needs to be cooled down slowly and uniformly, otherwise stresses form within its matrix leading to weakening and shattering at lower temperature.  For this reason, the temperature is reduced at $3\,\mathrm{^\circ C/min}$ from $800\,\mathrm{^\circ C}$ to $557\,\mathrm{^\circ C}$ which is the annealing point for borosilicate glass. At the annealing point stresses in the glass matrix can slowly relax \cite{eklund2007glass}. For our samples, $30\,\mathrm{mins}$ at $557\,\mathrm{^\circ C}$ is sufficient to reduce the stresses and eliminate the possibility of shattering as the glass returns to room temperature. The glass is further cooled at $2\,\mathrm{^\circ C/min}$ to $529\,\mathrm{^\circ C}$ which is the strain point of borosilicate D263 glass. Beyond the strain point the temperature can be safely dropped to room temperature at the maximum rate allowed by the furnace. Throughout the cooling process, the pressure in the furnace is maintained at $700\,\mathrm{mBar}$.

\begin{figure}
\centerline{\includegraphics[width=0.5\textwidth]{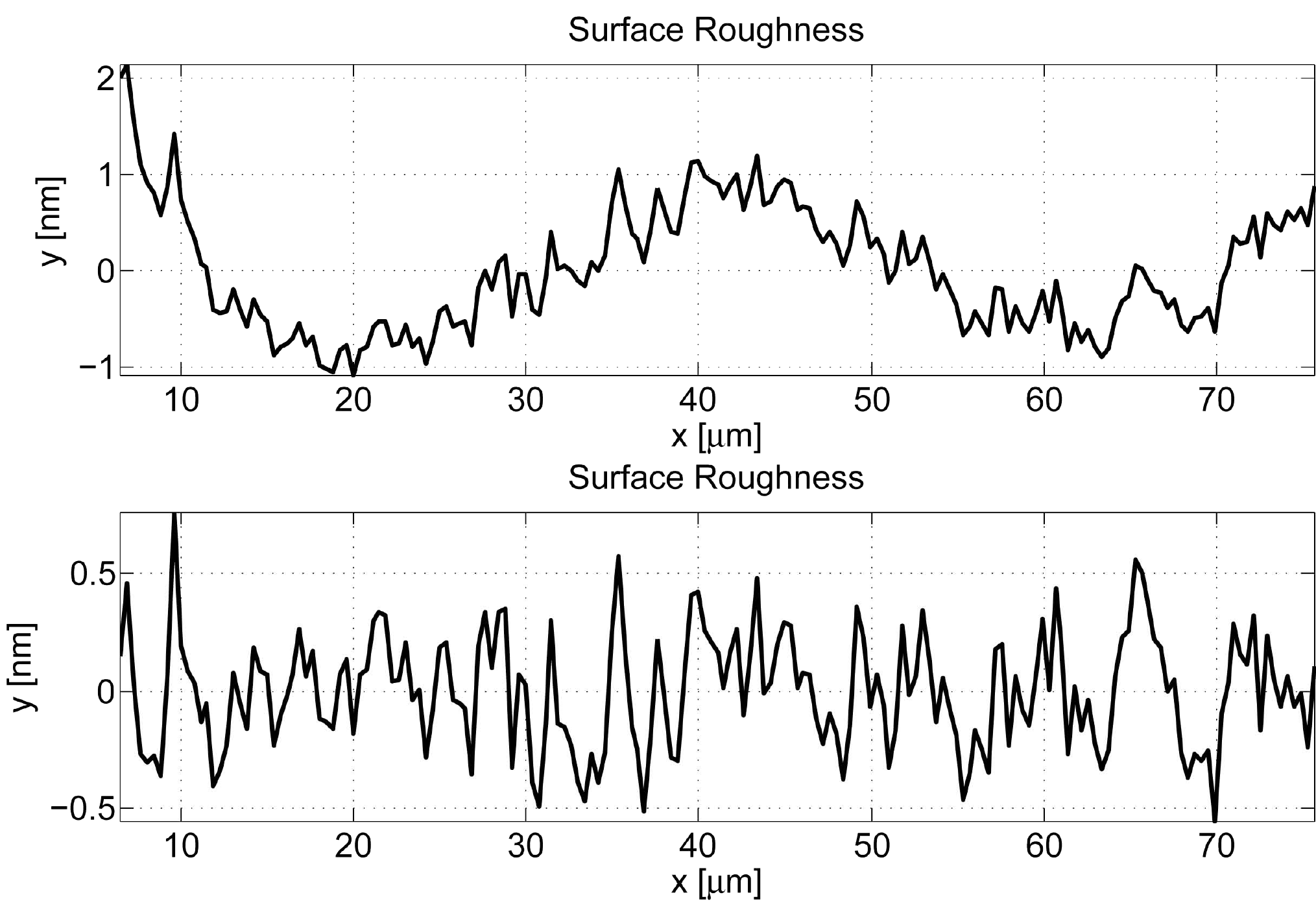}}
\caption{\label{surfaceprofile} Top (a) Surface roughness obtained after fitting the surface profile to a circle.
Below(b) Surface roughness obtained after fitting the surface profile to a $6\mathrm{th}$ order polynomial fit.}
\end{figure}

Mirrors were fabricated having a range of radius of curvatures ($0.5\,\mathrm{mm} - 5\,\mathrm{mm}$) made on MACOR substrates where the hole diameter varied from $1\,\mathrm{mm}$ to $3\,\mathrm{mm}$.  The surface profile was measured with an optical profiler that had a field of view of approximately $100\,\mathrm{\mu m}$. The resulting profile was first fitted to a circle and the residuals are given in Fig~\ref{surfaceprofile}(a).  The root-mean-square (rms) roughness calculated from these residuals is about 0.6 nm. However this is dominated by a large variation with a spatial period of about $50\,\mathrm{\mu m}$ clearly visible in the data.  We believe this to be an artifact of the optical profiler due to its presence in all test samples.  As this spatial period is not likely to impact on any practical application anyway, we use a $6\mathrm{th}$ order polynomial fit to better quantify the roughness and the residuals are given in Fig~\ref{surfaceprofile}(b).  The surface roughness determined from these residuals are typically $0.3\,\mathrm{nm}$ which is at the limits of the profiler sensitivity and comparable to best cavity mirror substrates available today.\cite{ATF}

Surface roughness results in scattering losses which limits the maximum possible reflectivity of the mirror \cite{scattering}.  When the surface roughness is much smaller then the wavelength of light, the scattering losses, $S$, are given by \cite{scattering}:
\begin{equation}
S = 1-\exp\left[- \left(\frac {4 \pi \sigma} {\lambda}\right)^2\right] \approx \left(\frac {4 \pi \sigma} {\lambda}\right)^2,
\label{reflectivityscattering}
\end{equation}
where $\sigma$ is the rms surface roughness and $\lambda$ is the wavelength of interest.  Using an operating wavelength of $\lambda=780\,\mathrm{nm}$, appropriate for $^{87}\mathrm{Rb}$ atoms, and our measured roughness of $\sigma=0.3\,\mathrm{nm}$ we infer that losses of $25\,\mathrm{ppm}$ or less should be possible with these substrates.

To better assess the suitability of the substrates for high finesse cavities, samples were coated with a high reflective coating and test cavities constructed. A target transmission of $100\,\mathrm{ppm}$ was chosen to ensure that scattering losses would not dominate the cavity losses resulting in negligible cavity transmission.  Four half symmetric cavities were constructed consisting of a flat mirror and a curved mirror with curvatures $0.7, 2, 5$ and $100\,\mathrm{mm}$.  The $100\,\mathrm{mm}$ curvature and flat mirrors were super-polished substrates available from the same company that provided the coating service \cite{ATF} and were coated in the same coating run as the custom substrates.  For each cavity we determine the cavity finesse, $\mathcal{F}$, and the single atom cooperativity, $\eta$. The finesse is a figure of merit for comparing the quality of different resonators and the single atom cooperativity quantifies the importance of cavity QED effects\cite{Vuletic}.  The parameters are given by the expressions\cite{Vuletic}
\begin{equation}
\label{parameters}
\mathcal{F}=\frac{c}{2 L \Delta \nu_c},\quad \eta = \frac{24 \mathcal{F}}{\pi k^2 w^2}
\end{equation}
where $\Delta \nu_c$ is the linewidth of the cavity (FWHM), $k=2\pi/\lambda$ and $w$ is the waist of the cavity mode which can be determined from the cavity geometry \cite{laser}. Thus, $\mathcal{F}$ and $\eta$ can be completely determined by the length of the cavity, the curvature of the mirrors, and the linewidth of the cavity.

The linewidth of the cavity is measured by modulating the laser with a broadband electro-optic modulator at a known frequency.  The resulting sidebands are then used to calibrate the frequency sweep as the laser is scanned across the resonance.  This provides a measurement of the cavity linewidth to better than $1\%$.

The radius of curvature is determined by maximizing the overlap of a focused beam with its reflection from the curved surface.  The output from a single mode fiber is collimated and then focussed by an aspheric lens to a waist of about $2\,\mathrm{\mu m}$.  The curved mirror is then used to reflect the light back through the fiber. The power reflected back through the fiber is maximized when the curvature of the laser wavefront and the mirror are equal.  This occurs at two points along the axis of a focused Gaussian beam and it is easily shown that these two points are separated by $R\sqrt{1-4 z_R^2/R^2}$ where $z_R$ is the Rayleigh length\cite{laser}.  This allows a measurement of $R$ which is accurate to about $20\,\mathrm{\mu m}$ for the tight curvature mirrors.  For the $100\,\mathrm{mm}$ curvature mirrors we simply use the value specified by the manufacturer.

The length of the cavity is measured using a high power optical microscope.  The calibrated focus adjustment allows us to measure the axial distance between the bottom of the mirror and a reference point on the mirror holder. Together with the holder dimensions we can then determine the length to an accuracy of about $25\,\mathrm{\mu m}$.

The measurements are summarized in Table~\ref{Cavity Data}.  Since the cavity finesse dropped sharply for the smaller curvature mirrors, we also measured their transmission directly.  The expected finesse is then calculated assuming that the transmission dominates the cavity losses.  The excellent agreement with the measured finesse and the values calculated from the measured transmission indicates that the finesse is completely dominated by the quality of the coatings and not the surface quality. The drop in the coating performance for small curvature mirrors is most likely due to shadowing effects arising from the large angular variation of the substrate surface and may be avoidable in future by further optimizing the coating run to specific curvatures\cite{Scot}.  However we note that, even with the degraded coating for tighter radius of curvature, the $\eta$ values indicate that the cavities are still suitable for cavity QED applications. We also note that the 2 and 5 mm curvature mirrors result in cavities which satisfy the stronger condition of strong coupling \cite{Vuletic}.  Moreover, since the cavity losses are dominated by transmission losses we can expect photons generated within the cavity to appear at the output with very high probability.

\begin{table}[h]
   \centering
    \begin{tabular}{|c|c|c|c|c|c|c|}
   \hline
 ROC & L&$T$ &  Finesse &Finesse&\\
(mm)&(mm)&(ppm)&(expected)&(obtained)& $\eta$\\ \hline
  100  & 2.075 &100 &31400$\pm$ 2000 &32200 $\pm$ 550 & 1\\
5  & 2.53&200 &22000 $\pm$ 2000 &19900 $\pm$ 340 & 3.8 \\
2 & 0.32 &350&13870 $\pm$ 1200 & 14200 $\pm$ 1460 & 9.2 \\
  0.7  & 0.25 &1500 &3850 $\pm$ 70 &3820 $\pm$ 500 & 5.4\\
  \hline
    \end{tabular}
    \caption{Summary of experimental cavities.  Expected finesse is calculated based on the measured transmission,T.  Length, L, and curvatures, RoC are measured as discussed in text.
\vspace{-5pt}
      \label{Cavity Data}}
\end{table}

In conclusion, we have successfully demonstrated the fabrication of spherical mirror substrates over a wide range of dimensions and demonstrated their potential application as Fabry Perot resonators in cavity QED applications. The dimensions of the mirrors produced, bridge the gap between existing commercially available mirrors and micro-mirrors based on silicon processing or fiber machining. This provides greater flexibility for incorporating optical resonators and light collection optics with micro-fabricated ion traps and atom chips. \cite{Isaac2, Blinov1}

\begin{acknowledgments}

We acknowledge technical assistance from Nick Lewty, Aarthi Dhanapaul, Jovan Kwek, and Kyle Arnold. This research was supported by the National Research Foundation and the Ministry of Education of Singapore, as well as by A-STAR under project No. SERC 052 123 0088.
\end{acknowledgments}


\begin{thebibliography}{25}
\bibitem [{ATF()}]{ATF}%
  \BibitemOpen
  \href@noop {} {}\bibinfo {note} {Superpolished mirrors and mirror coatings
  were provided by Advanced Thin Films.}\BibitemShut {Stop}%
\end{thebibliography}
\end{document}